\documentclass[12pt]{article}

\usepackage{amsmath}
\usepackage{amssymb}
\usepackage[dvips]{graphicx}
\usepackage{cite}

\makeatletter
 
  \@addtoreset{equation}{section}
 \makeatother

\newcommand {\n}{\nonumber \\}
\newcommand {\tr}{\mbox{tr}}

\begin{document}
\setlength{\oddsidemargin}{0cm}
\setlength{\baselineskip}{7mm}

\begin{titlepage}

~~\\

\vspace*{0cm}
    \begin{Large}
       \begin{center}
         {Four-algebraic extension of the IIB matrix model}
       \end{center}
    \end{Large}
\vspace{1cm}

\begin{center}
           Matsuo S{\sc ato}\footnote
           {
e-mail address : msato@cc.hirosaki-u.ac.jp}\\
      \vspace{1cm}
       
         {\it Department of Natural Science, Faculty of Education, Hirosaki University\\ 
 Bunkyo-cho 1, Hirosaki, Aomori 036-8560, Japan}

\end{center}

\hspace{5cm}

\begin{abstract}
\noindent

We make a four-algebraic extension of the IIB matrix model. The extension can be made by any Lie 4-algebra. The four-algebraic model has the same supersymmetry as the  IIB matrix model, and hence as type IIB superstring theory. The four-algebraic model contains twelve bosonic matrices; two of these will be identified with two extra dimensions that characterize F-theory. We construct a Lie 4-algebra that incorporates $u(N)$ Lie algebra and analyze the model explicitly by choosing it. We have three phases in the model with that specific algebra. In the first phase, it reduces to the original IIB matrix model. In the second phase, it reduces to a simple supersymmetric model. In the third phase, it reduces to a model that describes only the dynamics of the two matrices representing the torus.

\end{abstract}

\vfill
\end{titlepage}
\vfil\eject

\setcounter{footnote}{0}

\section{Introduction}
\setcounter{equation}{0}
The IIB matrix model was proposed as a non-perturbative formulation of type IIB superstring theory in 1996 \cite{IKKT} and has been extensively studied. It has the same space-time supersymmetry as the IIB superstring. The $\mathcal{N}=2$ supersymmetry in ten dimensions guarantees the existence of the graviton in this model.

In spite of the fact that the matrix model can describe some perturbative and non-perturbative dynamics of the IIB superstring, analyzing it is difficult as a consequence of its complicated interactions. New ideas are necessary to investigate non-perturbative dynamics of string theory. Extending matrix models for string theory can offer many ideas for studying non-perturbative dynamics of string theory, as well as the original matrix models. With this as motivation, extensions of the BFSS matrix theory and the IIB matrix model by Lie 3-algebra \cite{Nambu, Filippov, BergshoeffSezginTownsend, deWHN, Yoneya, Minic, Smolin1, Smolin2, Azuma, Nogo1, Jabbari, BLG1, Gustavsson, BLG2, Lorentz0, GomisSalimPasserini, HosomichiLeeLee, Nogo2, sp1, sp2, Nogo3, Lorentz1, Lorentz2, Lorentz3, Iso, ABJM, N=6BL, sp3, PangWang, IshiiIshikiShimasakiTsuchiya, ABJ, text, Nogo8, NishinoRajpoot, kac, bosonicM, Class, GustavssonRey, HanadaMannelliMatsuo, IshikiShimasakiTsuchiya, KawaiShimasakiTsuchiya, IshikiShimasakiTsuchiya2, MModel, DeBellisSaemannSzabo, Palmkvist, LorentzianM, ZariskiM} were studied in \cite{extBFSS, extIKKT}. The minimally extended models have two phases. They reduce to the original matrix models in one phase, whereas they reduce to simpler supersymmetric models in the other phase. The simple models are more tractable to investigate. 

In this paper, we further extend the IIB matrix model by using 4-algebras.  "Extend" indicates that the model is based on 4-algebras that incorporate Lie-algebras. The four-algebra model permits any 4-algebra whose quartic product is completely antisymmetric. We call such 4-algebra Lie 4-algebra. It has the same supersymmetry as the IIB matrix model, and hence as the IIB superstring.  As a consequence, the four-algebra model includes the graviton. The four-algebraic model contains twelve bosonic matrices; two of them will be identified with two extra dimensions that characterize F-theory.

We construct a Lie 4-algebra that incorporates $u(N)$ Lie algebra and analyze the model explicitly by choosing it. We have three phases in the model with that specific algebra. In the first phase, it reduces to the original IIB matrix model. In the second phase, it reduces to a simple supersymmetric model. In the third phase, it reduces to a model that describes only the dynamics of the two matrices representing the torus.

\vspace{1cm}

\section{Lorentzian Lie 4-algebra}

First, we construct a Lorentzian Lie 4-algebra. We consider an algebra
\begin{equation}
[T^A, T^B, T^C, T^D]= f^{A B C D}_{\qquad \,\,\, E} T^E,
\end{equation}
where the bracket is totally antisymmetric in A, B, C and D.
The gauge transformation is defined by
\begin{equation}
\delta X = \Lambda_{ABC} [T^A, T^B, T^C, X].
\end{equation}
The inverse of a metric is defined by
\begin{equation}
g^{AB}=<T^AT^B>.
\end{equation}
The indices $A, B, \cdots$ are raised and lowered by $g^{AB}$ and $g_{AB}$.
The gauge invariance of the metric 
\begin{equation}
<\delta T^A T^B>+ <T^A \delta T^B>=0
\end{equation}
requires
\begin{equation}
f^{ABCDE}=-f^{ABCED}.
\end{equation}
Then, the indices of $f^{ABCDE}$ are completely antisymmetric.
The fundamental identity is defined by
\begin{equation}
\delta [X, Y, Z, W]=[\delta X, Y, Z, W]+[X, \delta Y, Z, W]+[X, Y, \delta Z, W]+[X, Y, Z, \delta W],
\end{equation} 
which is equivalent to
\begin{equation}
f^{DEFG}_{\qquad \,\,\, H}f^{ABCHI}
=f^{ABCD}_{\qquad \,\,\, H}f^{HEFGI}
+f^{ABCE}_{\qquad \,\,\, H}f^{DHFGI}
+f^{ABCF}_{\qquad \,\,\, H}f^{DEHGI}
+f^{ABCG}_{\qquad \,\,\, H}f^{DEFHI}.
\end{equation}
We find a solution to this equation:
\begin{equation}
f^{\alpha \beta ijk}= G^{\alpha \beta} f^{ijk}
\end{equation}
with 
\begin{equation}
G^{\alpha \beta}= \kappa^{\alpha} t^{\beta} - t^{\alpha} \kappa^{\beta},
\end{equation}
where $\kappa^{\alpha}$ and $t^{\alpha}$ are arbitrary independent vectors and $f^{ij}_{\,\,\,\,\, k}$ are structure constants of ordinary Lie algebra. The other $f^{ABCDE}$ except for the above antisymmetrized form, are zero.
The non-zero metric is given by 
\begin{eqnarray}
&&g_{\alpha \bar{\beta}} = -\delta_{\alpha \beta} \n
&&g_{ij} = h_{ij},
\end{eqnarray}
where $h_{ij}$ is the Cartan metric of the Lie algebra.
Then, the non-zero commutators are
\begin{eqnarray}
&& [T^{\alpha}, T^{\beta}, T^i, T^j]=G^{\alpha \beta}[T^i, T^j] \n
&& [T^{\alpha}, T^i, T^j, T^k]=-f^{ijk}G^{\alpha}_{\,\,\, \bar{\beta}}T^{\bar{\beta}}.
\label{4-algebra}
\end{eqnarray}
This algebra includes the minimal Lorentzian Lie 3-algebra \cite{extBFSS, extIKKT} and thus an arbitrary Lie algebra.

\section{4-Algebraic Model}

In the present section, we construct a Lie 4-algebra model by extending the IIB matrix model. Our concerns are two scalars $\Phi^p$ $(p=1, 2)$, SO(1,9) vector $X^M$ $(M = 0, \cdots 9)$ and SO(1,9) Majorana-Weyl fermion $\Theta$ generated by Lie 4-algebra. $\Theta$ satisfies $\Gamma^{10}\Theta=-\Theta$. We do not presume specific algebra here. 

 The dynamical supertransformation of the IIB matrix model is extended by Lie 4-algebra:
\begin{eqnarray}
&&\delta X^{M} = i \bar{E} \Gamma^{M} \Theta \n
&&\delta \Phi^p = 0 \quad (p=1,2)\n
&&\delta \Theta = \frac{i}{2} [\Phi^1, \Phi^2, X_{M}, X_{N}] \Gamma^{MN} E,
\label{4algebraSUSY}
\end{eqnarray}
where $E$ satisfies
\begin{equation}
\Gamma^{10}E=-E.
\end{equation}
The algebra from this transformation is given by
\begin{eqnarray}
&&(\delta_2 \delta_1 - \delta_1 \delta_2)\Phi^p = 0 \n
&&(\delta_2 \delta_1 - \delta_1 \delta_2)X^{M} = 
-2 \bar{E}_2 \Gamma_{N} E_1 [\Phi^1, \Phi^2, X^{N}, X^{M}] \n
&&(\delta_2 \delta_1 - \delta_1 \delta_2)\Theta = 
-2 \bar{E_2} \Gamma_{N} E_1  [\Phi^1, \Phi^2, X^{N}, \Theta] \n
&& \qquad \qquad \qquad \qquad +(\frac{7}{8}\bar{E_2} \Gamma_{L} E_1 \Gamma^{L} -\frac{1}{8} \bar{E_2} \Gamma_{L_1 L_2 L_3 L_4 L_5} E_1 \Gamma^{L_1 L_2 L_3 L_4 L_5}) \n
&&  \qquad \qquad \qquad \qquad  \times
[\Phi^1, \Phi^2, X_{N} \Gamma^{N}, \Theta].
\label{4algebraSUSYalg}
\end{eqnarray}
The right-hand sides of the second and third lines imply the gauge transformation of $X^M$ and $\Theta$, individually.
The supersymmetry algebra closes on-shell if the fermion satisfies
\begin{equation}
[\Phi^1, \Phi^2, X_{N} \Gamma^{N}, \Theta]=0
\end{equation}
on-shell. If we transform this fermion equation of motion with (\ref{4algebraSUSY}), we get the boson equation of motion:
\begin{equation}
[\Phi^1, \Phi^2, X^{N}, [\Phi^1, \Phi^2, X_{N}, X^{M}]] - \frac{1}{2} [\Phi^1, \Phi^2, \bar{\Theta} \Gamma^{M}, \Theta] = 0.
\end{equation}
Both equations of motion can be obtained from the action
\begin{equation}
S=< -\frac{1}{4}[\Phi^1, \Phi^2, X^{M}, X^{N}]^2 
+\frac{1}{2}\bar{\Theta} \Gamma^{M} [\Phi^1, \Phi^2, X_{M}, \Theta] >.
\label{4algebraaction}
\end{equation}
This is invariant under (\ref{4algebraSUSY}).

The kinematical supersymmetry of the action (\ref{4algebraaction}) is generated by
\begin{equation}
\tilde{\delta} \Theta = \tilde{E}.
\end{equation}

The total supersymmetry algebra is 
\begin{eqnarray}
&&(\delta_2 \delta_1 - \delta_1 \delta_2)\Phi = 0 \n
&&(\delta_2 \delta_1 - \delta_1 \delta_2)X^{M} = 0\n
&&(\delta_2 \delta_1 - \delta_1 \delta_2)\Theta =0,
\end{eqnarray}
\begin{eqnarray}
&&(\tilde{\delta}_2 \tilde{\delta}_1- \tilde{\delta}_1 \tilde{\delta}_2) \Phi = 0 \n
&&(\tilde{\delta}_2 \tilde{\delta}_1- \tilde{\delta}_1 \tilde{\delta}_2) X^M =0 \n
&&(\tilde{\delta}_2 \tilde{\delta}_1- \tilde{\delta}_1 \tilde{\delta}_2) \Theta =0,
\end{eqnarray}
and
\begin{eqnarray}
&&(\tilde{\delta}_2 \delta_1- \delta_1 \tilde{\delta}_2) \Phi = 0 \n
&&(\tilde{\delta}_2 \delta_1- \delta_1 \tilde{\delta}_2) X^M = i \bar{E}_1 \Gamma^M E_2 \n
&&(\tilde{\delta}_2 \delta_1- \delta_1 \tilde{\delta}_2) \Theta = 0,
\end{eqnarray}
on-shell and up to the gauge symmetry.
If we change the basis as
\begin{eqnarray}
&&\delta' = \delta + \tilde{\delta} \n
&&\tilde{\delta}' = i(\delta - \tilde{\delta}),
\end{eqnarray}
we have
\begin{eqnarray}
&&(\delta_2' \delta_1'- \delta_1' \delta_2') \Phi = 0 \n
&&(\delta_2' \delta_1'- \delta_1' \delta_2')  X^M = i \bar{E}_1 \Gamma^M E_2 \n
&&(\delta_2' \delta_1'- \delta_1' \delta_2')  \Theta = 0,
\end{eqnarray}
\begin{eqnarray}
&&(\tilde{\delta}_2' \tilde{\delta}_1'- \tilde{\delta}_1' \tilde{\delta}_2') \Phi = 0 \n
&&(\tilde{\delta}_2' \tilde{\delta}_1'- \tilde{\delta}_1' \tilde{\delta}_2')  X^M = i \bar{E}_1 \Gamma^M E_2 \n
&&(\tilde{\delta}_2' \tilde{\delta}_1'- \tilde{\delta}_1' \tilde{\delta}_2')  \Theta = 0,
\end{eqnarray}
and
\begin{eqnarray}
&&(\tilde{\delta}_2' \delta_1'- \delta_1' \tilde{\delta}_2') \Phi = 0 \n
&&(\tilde{\delta}_2' \delta_1'- \delta_1' \tilde{\delta}_2') X^M = 0 \n
&&(\tilde{\delta}_2' \delta_1'- \delta_1' \tilde{\delta}_2') \Theta = 0.
\end{eqnarray}
This is the algebra of SO(1,9) $\mathcal{N}=2$ chiral supersymmetry, which is the supersymmetry algebra of the IIB matrix model as well as the IIB superstring. Therefore, this 4-algebraic model will be useful for studying F-theory by assuming that $\Phi^1$ and $\Phi^2$ represent the fixed torus that connects F-theory and IIB superstring theory.

\section{Model with A Certain 4-Algebra}

In the present section, we elucidate the Lie 4-algebra model with (\ref{4-algebra}) associated with $u(N)$.

The model allows BPS backgrounds
\begin{eqnarray}
&& \kappa^{\alpha}\Phi^1_{\alpha}=\bar{\Phi}^1, \quad
t^{\alpha}\Phi^1_{\alpha}=0 \n
&& \kappa^{\alpha}\Phi^2_{\alpha}=0, \quad 
t^{\alpha}\Phi^2_{\alpha}=\bar{\Phi}^2 \n
&& X^M_{\alpha}=\bar{X}^M_{\alpha} \n 
&& \Theta_{\alpha} = 0 \n
&& \Phi_i = X^M_i = \Theta_i = 0,
\label{BPSbg}
\end{eqnarray}
where $\bar{\Phi}^1$, $\bar{\Phi}^2$ and $\bar{X}^M_{\alpha}$ are arbitrary. Because the fluctuations of $\Phi^1_{\alpha}$, $\Phi^2_{\alpha}$, $X^M_{\alpha}$ and $\Theta_{\alpha}$ are zero modes around them, one needs to regard each of the backgrounds as independent vacuum and to fix the fluctuations.

The gauge transformation for an arbitrary field $X$,
\begin{equation}
\delta X_{\alpha} = \Lambda_{\beta \gamma \delta} f^{\beta \gamma \delta \eta}_{\quad \,\,\,\,\,\, \alpha} X_{\eta}
\end{equation}
can be explicitly written as
\begin{eqnarray}
&& \delta X_{\alpha} =0 \n
&& \delta X_i =  \Lambda^{(1) k}_{\quad \,\,\, i} X_k
+\Lambda_i^{(2)} \kappa^{\alpha} X_{\alpha}
+\Lambda_i^{(3)} t^{\alpha} X_{\alpha}.
\end{eqnarray}
There are three independent gauge parameters:
\begin{eqnarray}
&&\Lambda^{(1) k}_{\quad \,\,\, i}=3 G^{\alpha \beta} \Lambda_{\alpha \beta j} f^{jk}_{\quad i} \n
&&\Lambda_i^{(2)}= 3t^{\beta} \Lambda_{\beta jk} f^{jk}_{\quad i} \n
&&\Lambda_i^{(3)}= -3\kappa^{\beta} \Lambda_{\beta jk} f^{jk}_{\quad i},
\end{eqnarray}
where $\Lambda^{(1)}$ stands for the $u(N)$ transformation, while $\Lambda^{(2)}$ and $\Lambda^{(3)}$ stand for independent shift transformations.

In the $\bar{\Phi}^1 \neq 0$ case,
the shift transformation can fix one matrix as
\begin{equation}
\Phi^1= \Phi_{\alpha}^1 T^{\alpha} + \Phi_{\bar{\beta}}^1 T^{\bar{\beta}} + \Phi_i^1 T^i \to \Phi_{\alpha}^1 T^{\alpha} + \Phi_{\bar{\beta}}^1 T^{\bar{\beta}}. 
\end{equation}
Because $\Phi^1$ exist inside all the four-brackets in the action (\ref{4algebraaction}), non-zero four-brackets in the action reduce to three-brackets \cite{extBFSS, extIKKT} as
\begin{eqnarray}
&&[\Phi^1_{\alpha}T^{\alpha}, T^{\beta}, T^i, T^j]=\Phi^1_{\alpha}G^{\alpha \beta}[T^i, T^j] =\kappa^{'\beta}[T^i, T^j]=[T^{\beta}, T^i, T^j],\n
&& [\Phi^1_{\alpha}T^{\alpha}, T^i, T^j, T^k]=-\Phi^1_{\alpha}G^{\alpha}_{\,\,\, \bar{\beta}}f^{ijk}T^{\bar{\beta}}=-f^{ijk}\kappa^{'}_{\bar{\beta}}T^{{\bar{\beta}}}=[T^i, T^j, T^k],
\end{eqnarray}
where we redefine $\Phi^1_{\alpha}G^{\alpha \beta}$ as $\kappa^{'\beta}$. Then, the action (\ref{4algebraaction}) reduces to the minimally extended Lie 3-algebra IIB matrix model \cite{extIKKT},
\begin{equation}
S=< -\frac{1}{4}[\Phi^2, X^{M}, X^{N}]^2 
+\frac{1}{2}\bar{\Theta} \Gamma^{M} [\Phi^2, X_{M}, \Theta] >,
\label{extIIBaction}
\end{equation}
with the Lie 3-algebra 
 \begin{eqnarray}
&& [T^{\alpha}, T^i, T^j]=\kappa^{\alpha}[T^i, T^j] \n
&& [T^i, T^j, T^k]=-f^{ijk}\kappa_{\bar{\alpha}}T^{\bar{\alpha}}.
\end{eqnarray}
In the $\bar{\Phi}^1 \neq 0$ and $\bar{\Phi}^2 \neq 0$ case, one can fix one more matrix as
\begin{equation}
\Phi^2= \Phi_{\alpha}^2 T^{\alpha} + \Phi_{\bar{\beta}}^2 T^{\bar{\beta}} + \Phi_i^2 T^i \to \Phi_{\alpha}^2 T^{\alpha} + \Phi_{\bar{\beta}}^2 T^{\bar{\beta}}.
\end{equation}
Then, we get the IIB matrix model in a similar way:
\begin{equation}
S = \tr( -\frac{1}{4} [X^M, X^N]^2 
+ \frac{1}{2} \bar{\Theta} \Gamma^M[X_M, \Theta]). \label{IIBaction}
\end{equation}
This is consistent with the fact that the model (\ref{extIIBaction}) reduces to the IIB matrix model (\ref{IIBaction}) in the $\bar{\Phi}^2 \neq 0$ phase, as reported in \cite{extIKKT}.

In the $\bar{\Phi}^1 \neq 0$ and $\bar{\Phi}^2 = 0$ case, the model (\ref{extIIBaction}) reduces to the supersymmetric simple action,
\begin{equation}
S=\tr(-\frac{1}{2} [\Phi, X^{\mu}]^2
-\frac{1}{2} \bar{\Theta} \Gamma [\Phi, \Theta]),
\label{newaction}
\end{equation} 
as in \cite{extIKKT}. 

In the $\bar{\Phi}^1 =0$ and $\bar{\Phi}^2 = 0$ case, the action (\ref{4algebraaction}) reduces to
\begin{equation}
S=\mbox{tr}( -\frac{1}{4}(\bar{X}^{M}_{\alpha}\bar{X}^{N}_{\beta}G^{\alpha \beta}[\Phi^1, \Phi^2])^2 ).
\end{equation}
Without loss of generality, one can choose 
$\bar{X}^9_{\alpha}\kappa^{\alpha} \neq 0$, $\bar{X}^8_{\alpha}t^{\alpha} \neq 0$, and the other $\bar{X}^M_{\alpha}=0$. 
Because $\bar{X}^{M}_{\alpha}\bar{X}^{N}_{\beta}G^{\alpha \beta}=\bar{X}^{9}_{\alpha}\bar{X}^{8}_{\beta}G^{\alpha \beta}$ can be absorbed by redefinition of $\Phi^1$ and $\Phi^2$, this action is equivalent to
\begin{equation}
S= \mbox{tr}( -\frac{1}{4}[\Phi^1, \Phi^2]^2 ).
\label{torusaction}
\end{equation}
This is consistent with the supersymmetry transformation (\ref{4algebraSUSY}) because $\Phi^1$ and $\Phi^2$ are not transformed. In this phase, only the fields corresponding to the torus are dynamical.

%

\section{Conclusion and Discussion}
\setcounter{equation}{0}

In this paper, we have made a four-algebraic extension of the IIB matrix model. The extension can be made by any Lie 4-algebra. The four-algebraic model has the same supersymmetry as the  IIB matrix model, and hence as type IIB superstring theory. The four-algebraic model contains twelve bosonic matrices; two of these will be identified with two extra dimensions that characterize F-theory.

We have constructed a Lie 4-algebra that incorporates $u(N)$ Lie algebra and analyzed the model explicitly by choosing it. With that algebra, there are BPS moduli and we have gotten three phases. In the first phase, the model reduces to the original IIB matrix model. In the second phase, it reduces to the simple supersymmetric one (\ref{newaction}). In the third phase, it reduces to a model (\ref{torusaction}) that describes only the dynamics of the two matrices representing the torus. 

In this paper, we have constructed and chosen a specific Lie 4-algebra and studied the extended model explicitly, although we have constructed the 4-algebra model that allows any Lie 4-algebra. The next task is to construct and classify Lie 4-algebras as in \cite{Class} and apply them to the Lie 4-algebra model (\ref{4algebraaction}).

\vspace*{1cm}

\section*{Acknowledgements}
We would like to thank T. Asakawa, K. Hashimoto, N. Kamiya, H. Kunitomo, T. Matsuo, S. Moriyama, K. Murakami, J. Nishimura, S. Sasa, F. Sugino, T. Tada, S. Terashima, S. Watamura, K. Yoshida, and especially H. Kawai and A. Tsuchiya for valuable discussions. This work is supported in part by Grant-in-Aid for Young Scientists (B) No. 25800122 from JSPS.

\end{document}